\documentclass[final,3p,times,twocolumn]{elsarticle}

\usepackage{graphics}

\usepackage{amssymb}

\journal{Physica C}

\begin{document}

\begin{frontmatter}



\title{Effect of stripe order strength for the Nernst effect in La$_{2-x}$Sr$_x$CuO$_4$ single crystals}

\author[label1]{T. Fujii}
\ead[label1]{fujii@crc.u-tokyo.ac.jp}
\author[label2]{T. Matsushima}
\author[label2]{T. Maruoka}
\author[label1,label2]{A. Asamitsu}
\address[label1]{Cryogenic Research Center, University of Tokyo, Tokyo, 113-0032, Japan}
\address[label2]{Department of Applied Physics, University of Tokyo, Tokyo, 113-8656, Japan}

\begin{abstract}
We have precisely measured the Nernst effect in Nd-doped La$_{2-x}$Sr$_x$CuO$_4$ single crystals with controlling the strength (stability) of the stripe order.
We found that the onset temperature $T_{onset}$, where the Nernst signal starts increasing, does not change 
conspicuously in spite of Nd-doping. 
At low temperatures, on the other hand, the absolute value of the Nernst signal is strongly suppressed in accordance with the strength of the stripe order. 
These results imply that the fluctuation of (charge) stripe order enhances the Nernst signal below $T_{onset}$ at high temperatures,
and then the stripe order enhanced by Nd-doping suppresses the superconducting fluctuation to reduce the Nernst signal at low temperatures.  
We also observed an increase of the Nernst signal below the charge order temperature $T_{ch}$ which is observed in diffraction measurement.
\end{abstract}


\end{frontmatter}


\section{Introduction}
\label{}
The Nernst effect in the normal state of high-$T_c$ cuprates has become a focus of renewed attention since the detection of an anomalous Nernst signal 
in the normal state of underdoped cuprates~\cite{natureong}. 
It is widely accepted that the large Nernst signal is attributed to the movement of vortices which survives far above $T_c$. 
On the other hand, a large negative Nernst coefficient, which is comparable to the vortex signal in the superconducting state, has been reported in the charge 
density wave state of NbSe$_2$~\cite{Bel}. 
Recently, it is reported that the stripe order causes the Nernst signal to become large in Nd or Eu doped La$_{2-x}$Sr$_x$CuO$_4$~\cite{natureLSCO}. 
Thus, two mechanisms for the origin of Nernst effect have been discussed up to now. 
The former supports the scenario that the superconducting gap opens far above $T_c$, although the long-range phase coherence is destroyed by vortex excitation.
While the latter implies that the superconductivity occurs near the quantum critical point where the competing orders, such as stripe order and/or charge order, 
cross in the phase diagram. 
Here, to investigate the effects of the charge (stripe) ordered state on Nernst signal, we controlled the strength (stability) of the stripe order 
by introducing Nd in La$_{2-x}$Sr$_x$CuO$_4$, and measured the Nernst effect.

\section{Experimental}
The high-quality single crystals of La$_{1.85-y}$Nd$_y$Sr$_{0.15}$CuO$_4$ (y=0, 0.2, and 0.4) were grown using the Traveling Solvent Floating Zone method. 
Grown boule were carefully cut into rectangular shape confirming the crystal axis by using X-ray laue camera. 
All samples were carefully prepared to have the same doping level by fixing the Sr content (x=0.15) and the doping level was confirmed by 
the absolute values of the thermopower at room temperature.  $T_c$'s determined by susceptibility were 37K, 23K, and 12K for y=0, 0.2, and 0.4 respectively.
For the samples of y=0.2 and 0.4, we observed an upturn in the resistivities around 50K and 70K, respectively, at which charge order has been observed~\cite{phasediagram}.
Nernst signal $e_y$, which is defined as the transverse electric field $E_y$ per unit temperature-gradient ($e_y(H,T)=E_y/\nabla T$), was measured with applying three 
different temperature-gradients ($\nabla T \approx$ 0.5, 0.75, 1K/mm). Temperatures and magnetic fields were stabilized during the measurement.

\section{Results and Discussion}
Since the Nernst signal at high temperature is linear to magnetic field, we can define the Nernst coefficient $\nu$ as $e_y(H,T)/B$, where $B=\mu_0H$. 
Figure 1 shows the temperature dependence of $\nu$ in high temperature. To clarify the doping dependence, $\nu$ of Nd-free sample with $x$=0.125 is also plotted. 
Note that the full scale of the vertical axis is as small as 20 nV/KT, which assures that the resolution of the measurement is better than 5 nV/KT. 
As seen in the figure, $\nu$'s of x=0.15 are completely coincide irrespective of Nd content. The magnitude of $\nu$ is almost consistent with the previous report~\cite{natureong}. 
These results suggest that the absolute value of $\nu$ at high temperatures is determined only by the doping level $x$.
More interestingly, the onset temperature $T_{onset}$, below which $\nu$ starts to deviate from high-temperature $T$-linear behavior (normal-term Nernst coefficient), 
does not change with Nd-doping, even though the superconducting transition temperature $T_c$ is drastically reduced. 
\begin{figure}
\begin{center}
\includegraphics[width=6.8cm,clip]{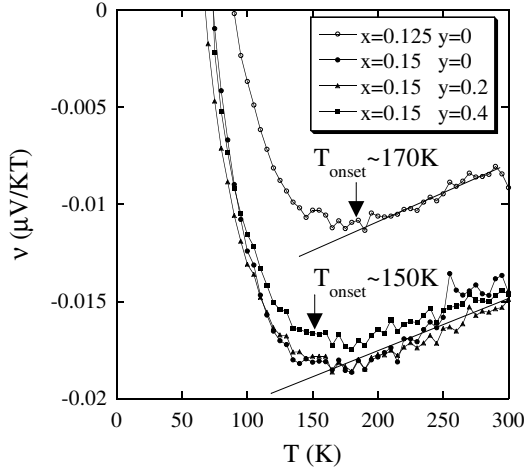}
\end{center}
\caption{Nernst coefficient $\nu$ of La$_{2-x-y}$Nd$_y$Sr$_{x}$CuO$_4$ in high temperature.}
\label{f1}
\end{figure}

To investigate the effect of the stripe order, we show the temperature dependence of $e_y$ at 9T in Fig. 2. The magnitude of 
$e_y$ drastically decreases with Nd-doping below the temperature $T_B$, where the magnetic field dependence of the Nernst coefficient was observed ($T_B$ was determined as ref.[3]). 
This result indicates that the fluctuation of superconductivity is suppressed by the stabilization of the stripe order due to Nd-doping below $T_B$.
As seen in La$_{1.8-x}$Eu$_{0.2}$Sr$_x$CuO$_4$ of ref.[3], we observed the evolution of two-peak structure in the sample of x=0.4. 
Furthermore, an upturn of the Nernst signal has been clearly observed below the charge order temperature $T_{ch}$ (see the inset of Fig.2).
This is crucial evidence that the static stripe order enhances the Nernst signal.

Based on the above results, we can define three characteristic temperatures: $T_{onset} > T_{ch}>  T_B$.
With decreasing temperature, the following scenario is expected at this stage:   
On the assumption that the enhancement of the Nernst signal is attributed to the effect of the stripe order at high temperatures, 
the fluctuation of the stripe order develops below $T_{onset}$, and then the static stripe order enhances the Nernst signal further at $T_{ch}$.
With further decrease of temperature, a huge Nernst signal induced by the vortex flow is expected below $T_B$.


\begin{figure}
\begin{center}
\includegraphics[width=6.3cm,clip]{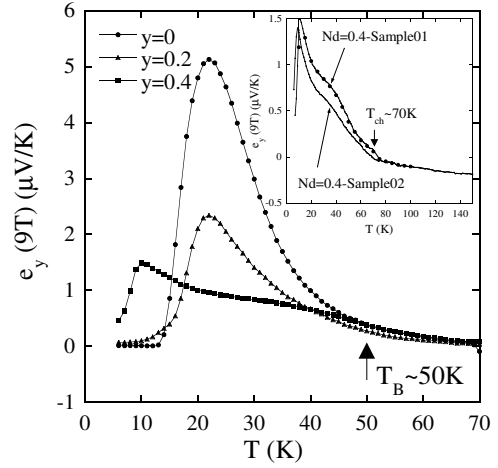}
\end{center}
\caption{Nernst signal e$_y$ at 9T for La$_{1.85-y}$Nd$_y$Sr$_{0.15}$CuO$_4$. $T_B$ was determined as ref.[3].
Inset: Nernst signal e$_y$ at 9T for y=0.4. To confirm reproducibility, we measured two different sample.}
\label{f2}
\end{figure}

\section{Conclusion}
In order to investigate the effect of the charge (stripe) order on Nernst signal, we measured the temperature and magnetic-field dependences of the Nernst effect 
in Nd-doped La$_{2-x}$Sr$_x$CuO$_4$ single crystals with controlling the strength (stability) of the charge order.
We found that the magnitude of Nernst signal drastically decrease with Nd doping below $T_B$, although 
$T_{onset}$ does not change conspicuously, and that the Nernst signal is also enhanced at the charge ordered transition temperature $T_ch$.
These results suggest that the stripe order as well as its fluctuation enhances the Nernst signal at high temperatures, whereas
the stabilized stripe order tends to suppress the superconducting fluctuation and hence to reduce the Nernst signal at low temperatures. 

\end{document}